\documentclass[aps,pra,linenumber]{revtex4}
\usepackage{graphicx}
\usepackage{dcolumn}
\usepackage{bm}
\usepackage{amsmath}
\usepackage{amssymb,amsthm}
\usepackage{epsfig,color}
\usepackage[sans]{dsfont}
\usepackage{multirow}
\usepackage{lineno}
\usepackage{graphicx}

\allowdisplaybreaks

\definecolor{nblue}{rgb}{0.2,0.2,0.7}
\definecolor{ngreen}{rgb}{0.2,0.6,0.2}
\definecolor{nred}{rgb}{0.7,0.2,0.2}
\definecolor{nblack}{rgb}{0,0,0}

\def\bea{\begin{eqnarray}}
\def\eea{\end{eqnarray}}

\begin{document}

\title{Quantum Fisher information of fermionic cavity modes in an accelerated motion}%

\author{Zahid Hussain Shamsi, Dai-Gyoung Kim}

\affiliation{Department of Applied Mathematics, Hanyang University
(ERICA), Ansan, Korea 425-791}

\author{Younghun Kwon}
\email{yyhkwon@hanyang.ac.kr}

\affiliation{Department of Applied Physics, Hanyang University (ERICA), Ansan, Korea 425-791}


\begin{abstract}
We investigate the effect of the inertial and non-inertial segments
of relativistic motion on the quantum Fisher information of
$\left(1+1\right)$  Dirac field modes confined to cavities. For the
purpose we consider $\theta$ parameterized two-qubit pure entangled
state. In this setting, the initial state is obtained by a unitary
operation which entangles the two confined modes of cavities of
Alice and Bob. In the situation that Rob's cavity, initially
inertial, accelerates uniformly with respect to its proper time and
then again becomes inertial while Alice's cavity remains inertial, 
we analyze the quantum Fisher
information of the system. First of all, we find that the quantum
Fisher information of the pure composite system $\mathcal{F}_\theta$
with respect to parameter $\theta$ is invariant regardless of the
non-inertial movement. However in the quantum Fisher information
distribution over the subsystems of Alice's and Rob's cavities, the
quantum Fisher information over the Rob's cavity is shown to be the
periodic degradation behavior depending upon the parameter $\theta$.
In order to check whether the invariance of the quantum Fisher
information of the composite system is an intrinsic property, we
consider a depolarizing state of the pure composite system. Then we
show that the quantum Fisher information of the Werner state with
respect to parameter $\theta$ is not invariant. Furthermore for the
non-inertial motion of a single cavity, we find that  the quantum
Fisher information with respect to parameter $\theta$ for a $\theta$
parameterized initial state is degraded in terms of acceleration.
Finally subadditivity of the quantum Fisher information is shown
regardless of the pure composite system or the Werner state.
\end{abstract}
\maketitle

\section{%
Introduction}\label{sec:intro} Recently quantum Fisher information
(QFI) \cite{Braunstein1996135,PhysRevLett.72.3439,1976quantum} has
gained considerable attention as a promising candidate to understand
the information contents of quantum states. It is well known that
QFI has been successfully applied to quantum statistical inference
and estimation theory
\cite{PhysRevLett.91.180403,PhysRevA.84.042121}. When information is
encrypted in a set of quantum states, more often in terms of
appropriate parameters, extraction of the encoded information
requires the distinction between  different quantum states. This
requirement can be accomplished through measurements that yield a
set of observations characterized by those parameters. Therefore the
problem of distinguishing the quantum states is transformed into a
problem of parameter estimation.\\
\indent The classical Fisher information is known to provide the
lower bound on the variance of the parameter estimation in terms of
the well known Crame$\acute{}$r-Rao bound. For asymptotically large
sets of observations, the maximum likelihood estimation (MLE)
approach provides unbiased parameter estimation that can attain this
bound \cite{Fisher1925}. However, when one should deal with quantum
information, since quantum information content is much more complex
and intrinsically contains non-trivial characteristics that can not
be associated with their classical counterpart
\cite{Braunstein1996135,holevo2011probabilistic}, the quantum Fisher
information should be coined
\cite{1976quantum,holevo2011probabilistic,PhysRevLett.72.3439} as a
natural extension to the classical Fisher information.
\par Furthermore, successful efforts to combine
relativity and quantum information theory have opened new avenues to
explain quantum behavior at a macroscopic scale
\cite{PhysRevD.14.2460,PhysRevD.14.870,Wald94quantumfield,fabbri2005modeling,RevModPhys.76.93}.
A focus of this area of research is the dynamics of information
contents under the Unhruh-Hawking effect
\cite{AlcingObDEpEnt-0264-9381-29-22-224001,PhysRevA.82.042332,RevModPhys.80.787}.
For example, the dynamics of entanglement
\cite{PhysRevLett.95.120404,PhysRevD.85.061701,PhysRevA.87.022338,PhysRevD.85.025012,PhysRevA.74.032326},
quantum discord
\cite{PhysRevA.80.052304,PhysRevA.81.052120,PhysRevA.86.032108}, and
fidelity of teleportation
\cite{PhysRevLett.91.180404,PhysRevLett.110.113602} have already
been studied. For instance, the correlations induced by maximally
entangled bosonic or fermionic bipartite states were found to be
degraded due to the accelerated motion of one party with respect to
the other inertial party
\cite{PhysRevD.85.061701,PhysRevD.85.025012}. \\
 \indent The study of Unruh-Hawking effect  on the quantum Fisher
information is a crucial step to use QFI as a reliable resource for
quantum metrology in a relativistic regime. Aspachs et al. described
the optimal detection process of Unruh-Hawking effect itself and
showed that the Fock states can achieve maximal QFI with a scalar
field in a two-dimensional Minkowski spacetime
\cite{PhysRevLett.105.151301}. In \cite{Yao2014} 
authors investigated the performance of QFI for both scalar and
Dirac fields parameterized by weight and phase parameters in
non-inertial frames. In \cite{MehdiAhmadi2013,PhysRevD.89.065028},
the authors evaluated the QFI of the acceleration parameter for the
relativistic scalar and Bosonic quantum  fields confined to a cavity
using Bures Statistical distance \cite{BuresMR0236719}.
\par The main contribution of this paper is to investigate QFI under the influence of non-uniform motion,
on the precision of parameter estimation for the parameter $ \theta
$. The parameter $ \theta $ is encoded through a unitary operation,
which causes the two-qubit entangled state of the Dirac fields
confined in two cavities. In other words, the entangling power of a
unitary operation is precisely unknown, and the parameter $ \theta $
arises naturally. Although, the evaluation of QFI for various
physical parameters in the inertial frames is equally important and
has been studied recently in different scenarios, in this manuscript
the QFI dynamics is investigated in the perspective of non-uniform
motion.
\par More precisely, we follow the Dirac field analysis proposed in [19],
where the modes of massless Dirac field are confined to the two
cavities with Dirichlet's boundary conditions and one of the
cavities remains inertial while the other cavity undergoes the
segments of inertial and non-inertial motion with uniform
acceleration. We restrict the uniform acceleration to be very small
$\left(a\ll 1\right)$ and use perturbation theory to observe the
variation of the QFI for the confined modes of the Dirac field with
respect to the parameter $\theta$, under the accelerated motion.
\par The rest of the paper is organized as follows. In Sec.~II, 
we briefly review the properties of QFI and discuss the recent developments
regarding its analytic evaluation. The perturbed Bogoliubov
coefficients and Fock space quantization for vacuum and one charged particle fermionic states in the cavities are described in  Sec.~III.
The QFI for two mode fermionic Fock state shared between Alice's and
Rob's cavities and distribution of QFI over each cavity is discussed
in Sec.~IV. And the behavior of QFI for parameterized initial state
in a single cavity under the non-initial motion is studied.
In Sec.~V,
we extend our investigation to Werner state \cite{PhysRevA.40.4277}. In Sec.~VI,
we conclude and discuss our results.
\section{%
Quantum Fisher Information}\label{sec:2QFI_review}
 Consider a quantum state $\rho_\lambda$ having a one parameter family of
N-dimensional quantum states, with parameter $\lambda$. A set of
quantum measurements $\{E\left(\xi\right)\}$, which is POVM, should
be performed on $\rho_\lambda$ to extract the information about
unknown parameter $\lambda$. This process resembles the classical
case where the parameter estimation is performed from the observed
data obtained from classical experiments and the efficiency of the
estimate(s) is measured in terms of the classical Fisher information
\cite{1976quantum,Fisher1925,holevo2011probabilistic}
 \begin{equation}\label{eq:FisherInfoClassic}
 F_\lambda=\int d\xi p\left(\xi|\lambda\right)
\left[ \partial_\lambda \ln p\left(\xi|\lambda\right)\right]^2=
 \int d\xi
 \frac{\left(\partial_\lambda p\left(\xi|\lambda\right)\right)^2}{p\left(\xi|\lambda\right)},
 \end{equation}
 where $p\left(\xi|\lambda\right)$ denotes the conditional probability of observing the result $\xi$ provided the value of the parameter is $\lambda$.
However, the quantum Fisher information differs significantly in the
operational sense due the fact that the probability
$p\left(\xi|\lambda\right)$ results from measurement by the quantum
operator $E\left(\xi\right)$ rather than a mere classical
experiment.
 According to Born's rule,  $p\left(\xi|\lambda\right)=Tr\left[E\left(\xi\right)\rho_\lambda\right]$ represents
the conditional probability of obtaining the outcome $\xi$  when the
given value of the parameter is $\lambda$.  In the quantum world,
the parameter estimation problem can be thought of as the problem of
searching for the set of measurements $\{E\left(\xi\right)\}$ that
yields the optimal value of (\ref{eq:FisherInfoClassic}). Therefore,
the quantum Fisher information can be defined as
\cite{Braunstein1996135,PhysRevLett.72.3439}
\begin{equation}\label{eq:QauntumFI}
\mathcal{F}_\lambda=\max_{E\left(\xi\right)}F_\lambda.
\end{equation}
Further, using the notion of symmetric logarithmic derivative (SLD) introduced in \cite{1976quantum}
\begin{equation}\label{eq:SymmLogarithmicDerivativeQuantum}
 \partial_\lambda \rho_\lambda=\frac{1}{2}\left(\rho_\lambda L_\lambda+L_\lambda \rho_\lambda\right),
\end{equation}
we obtain an alternate expression for the QFI as
\begin{equation}\label{eq:QauntumFIrhoL2form}
\mathcal{F}_\lambda=Tr\left(\rho_\lambda L_\lambda^2\right)=Tr\left(\partial_\lambda\rho_\lambda L_\lambda\right),
\end{equation}
where the operator $L_\lambda$ is the solution of the Lyapunov
matrix equation \cite{PARIS2009} given by
Eq.~(\ref{eq:SymmLogarithmicDerivativeQuantum}). It is worth
noticing that expressions of QFI in (\ref{eq:QauntumFI}) and
(\ref{eq:QauntumFIrhoL2form}) are equivalent in the sense that the
operator $L_\lambda$ admits the spectral decomposition in the same
basis as used for the spectral decomposition of $\rho_\lambda$.
 Therefore, based on the spectral decomposition $\rho_\lambda=\sum_{i=1}^N p_i|\psi_i\times\psi_i|$, the operator $L_\lambda$ is written as \cite{PARIS2009}
 \begin{equation}\label{eq:LogDerivavtivspectralForm}
 L_\lambda=2\sum_{m,n}^{N}\frac{\left|<\psi_m\left|\partial_\lambda\rho_\lambda\right|\psi_n>\right|}{p_m+p_n}\left|\psi_m\times\psi _n\right|,
\end{equation}
where the eigenvalues $p_i\ge0$ and $\sum_{i=1}^N p_i=1$. In turn,
the set of eigenvectors of  $L_\lambda$ provides the optimal POVM to
obtain the QFI. Making use of (\ref{eq:QauntumFIrhoL2form}) and
(\ref{eq:LogDerivavtivspectralForm}), the QFI can be rewritten as
\cite{PARIS2009}
\begin{equation}\label{eq:QauntumFI_Spectralform}
\mathcal{F}_\lambda=2\sum_{m,n}^{N}\frac{\left|<\psi_m\left|\partial_\lambda\rho_\lambda\right|\psi_n>\right|^2}{p_m+p_n}.
\end{equation}
Using the completeness relation
\begin{equation}\label{eq:completeness}
\sum_{i=M+1}^{N}|\psi_i\times\psi_i|=\mathbb{I}-\sum_{i=1}^{M}|\psi_i\times\psi_i|,
\end{equation}
where M denotes the number of eigenvectors corresponding to
non-zero eigenvalues and represents the dimension of the support of
$\rho_\lambda$, Zhang et al. provided a relatively simple
expression to evaluate the QFI for general states as
\cite{PhysRevA.88.043832}
\begin{equation}\label{eq:QauntumFIDesiredForm}
\mathcal{F}_\lambda=\!\!\sum_{i=1}^M\frac{\left(\partial_\lambda p_i\right)^2}{p_i}+\!\!\sum_{i=1}^M4p_i\mathcal{F}_{\lambda,i}
-\!\!\sum_{i\neq j}^M\frac{8p_ip_j\left|\big<\psi_i|\partial_\lambda\psi_j\big>\right|^2}{p_i+p_j},
\end{equation}
where $\mathcal{F}_{\lambda ,i}$ is the QFI of the pure state $|\psi_i>$ given by
\begin{equation}\label{eq:QFIPureState}
\mathcal{F}_{\lambda,i}
=\big<\partial_\lambda\psi_i|\partial_\lambda\psi_i\big>-\left|\big<\psi_i|\partial_\lambda\psi_i\big>\right|^2.
\end{equation}
It is noteworthy that the QFI of a low rank density matrix is now
only determined by its support spanned by the eigenvectors,
$|\psi_i\big>; \quad i=1,\cdots, M $, corresponds to nonzero
eigenvalues. In fact, the first term on the right hand side of
(\ref{eq:QauntumFIDesiredForm}) represents the classical case where
the set of nonzero eigenvalues behaves like a classical probability
distribution. And the second contribution of
(\ref{eq:QauntumFIDesiredForm}) denotes the weighted average of QFI
due to all the pure-states. The last term of
(\ref{eq:QauntumFIDesiredForm}) captures the QFI from the mixture of
pure states and therefore causes the decrease in the total QFI.
Therefore, the last two terms can be interpreted as the quantum part
of the Fisher information while the first term represents its
classical constituent. In this paper, we will exploit Eq.
(\ref{eq:QauntumFIDesiredForm}) due to its relative simplicity and
clear physical interpretation to find the QFI for the modes of the
Dirac field confined to the inertial and non-inertial cavities.
\section{%
Bogoliubov Transformation For Inertial and Non-inertial
Segments}\label{sec:4Pert_Bogoliubov}
In order to find the unitary
transformation of cavity's transitions between the inertial and
non-inertial segments of motion, we setup two cavities with the
observers referred to as  Alice and Rob, respectively. Both the
cavities are inertial and completely overlap at $t=0$.
Here we assume that the cavity walls are placed at $x = a$ and $x = b$
where $0 < a < b$. Rob's cavity
then moves with uniform acceleration to the right along the
time-like killing vector $\partial_\eta$ for duration $\eta=0$ to
$\eta=\eta_1$  in the Rindler co-ordinates. Duration of the
acceleration, $\frac{2}{a+b}$, with respect to proper time measured
at the center of the cavity is thus $\tau_1=\frac{a+b}{2}\eta_1$.
Finally, Rob's cavity again becomes inertial with respect to its
rest frame. The Alice's cavity remains inertial throughout this trip
of Rob's cavity.
Therefore, three segments of Rob's trajectories can be identified as Regions I, II and III.\\
This grafting process of cavity motion is explained here to make this paper sufficiently self contained.
Earlier the same process is introduced and exploited by \cite{ARLEEPHDThesis,PhysRevD.85.025012} to study the dynamics of the entanglement for bosonic and fermionic cavities.\\
The Dirac field representation in three regions is
\begin{subequations}
\begin{eqnarray}
\label{eq:DiracFieldRepresntation1}
\textup{I:}\quad \psi&=&\sum_{n\ge0}a_n\psi_n+\sum_{n<0}b_n^\dag\psi_n,
\\*\label{eq:DiracFieldRepresntation2}
\textup{II:}\quad \psi&=&\sum_{n\ge0}\hat{a}_n\hat{\psi}_n+\sum_{n<0}\hat{b}_n^\dag\hat{\psi}_n,
\\*\label{eq:DiracFieldRepresntation3}
\textup{III:} \quad \psi&=&\sum_{n\ge0}\tilde{a}_n\tilde{\psi}_n+\sum_{n<0}\tilde{b}_n^\dag\tilde{\psi}_n,
\end{eqnarray}\label{eq:DiracFieldRepresntation123}
\end{subequations}
with the respective non vanishing anticommutators
\begin{subequations}
\begin{eqnarray}\label{eq:DiracFieldAntiCommutationRelation123}
\textup{I:}\quad \{a_m,a_n^\dag\}&=&\{b_m,b_n^\dag\}=\delta_{mn},\\*
\textup{II:}\quad \{\hat{a}_m,\hat{a}_n^\dag\}&=&\{\hat{b}_m,\hat{b}_n^\dag\}=\delta_{mn},\\*
\textup{III:}\quad \{\tilde{a}_m,\tilde{a}_n^\dag\}&=&\{\tilde{b}_m,\tilde{b}_n^\dag\}=\delta_{mn}.
\end{eqnarray}
\end{subequations}
Using Bogoliubov transformation, the Dirac field modes between Region I and II are related as \cite{PhysRevD.85.025012}
\begin{equation}\label{eq:BogTI_II}
\hat{\psi}_m=\sum_n A_{mn}\psi_n,
\end{equation}
where ${\psi}_n$ and $\hat{\psi}_m$ are the Dirac field modes in
regions I and II respectively. For the small acceleration case,
Friis et al. \cite{PhysRevD.85.025012} derived these coefficients
$A_{mn}$ in the perturbative regime by introducing the dimensionless
parameter $h=\frac{2L}{a+b}$, satisfying $0<h<2$. These coefficients
preserve the unitarity of the transformation to the order
$O\left(h^2\right)$ for $s>0, s\rightarrow0_+$ and are given in
terms of Maclaurin's series expansion as
\cite{PhysRevD.85.025012,ARLEEPHDThesis}
\begin{equation}\label{eq:Bogliubov_II_MacSerieExp}
A_{mn}=A_{mn}^{\left(0\right)}+A_{mn}^{\left(1\right)}+A_{mn}^{\left(2\right)}+O(h^3),
\end{equation}
where the superscripts indicate the power of parameter $h$. During
the non-inertial trajectory, modes $\hat{\psi}_m$ in Rob's cavity
remain independent and do not interact. Hence, these modes can only
develop some phases during the non-inertial duration
$0\le\eta\le\eta_1$. This change in the modes can be balanced by
introducing a diagonal matrix $G\left(\eta_1\right)$ whose diagonal
entries are \cite{PhysRevD.85.025012}
\begin{equation}\label{eq:PhaseDeveloped_RII}
G_{nn}\left(\eta_1\right)=\exp\left(i\Omega_n\eta_1\right).
\end{equation}
 For $\eta \ge \eta_1$, the transformation from region II to region III can be obtained by simply using the inverse transformation $A^\dag=A^{-1}$.
 The evolution of the Dirac field mode from region I to region III in Rob's cavity can then be expressed by the Bogoliubov transformation matrix
 \begin{equation}\label{eq:TotalMatrixCompsition}
\mathcal{A}=A^\dag G\left(\eta_1\right) A.
 \end{equation}
 Thus the Bogoliubov transformation for the Dirac field modes between regions I and III reads
 \begin{equation}\label{eq:ModeRelation13}
\tilde{\psi}_m=\sum_n\mathcal{A}_{mn}\psi_n.
 \end{equation}
 It can be noticed that $\mathcal{A}$, being the composition of unitary matrices,  is also a unitary matrix to the order $h^2$. Similarly, the Bogoliubov transformation for the Dirac field mode
operators can also be expressed as \cite{ARLEEPHDThesis,PhysRevD.85.025012}
 \begin{subequations}
 \begin{eqnarray}
 k>0:\qquad a_k&=&\sum_{l\ge0}\tilde{a}_l\mathcal{A}_{lk}
 +\sum_{l<0}\tilde{b}_l^\dag\mathcal{A}_{lk},\\*
  k<0:\qquad b_k&=&\sum_{l\ge0}\tilde{a}_l^\dag\mathcal{A}_{lk}
 +\sum_{l<0}\tilde{b}_l\mathcal{A}_{lk}.
 \end{eqnarray}\label{eq:ModeOpoeratorsRelation13}
  \end{subequations}
The relation between the Fock vacua in regions I and III denoted by
 $|0\big>$ and $|\tilde{0}\big>$ is \cite{fabbri2005modeling,PhysRevD.85.025012}
 \begin{equation}\label{eq:vacuRegion13}
 |0\big>=Ne^W|\tilde{0}\big>,
 \end{equation}
 where
 \begin{equation}\label{eq:W}
W=\sum_{p\ge0, q<0}V_{pq}\tilde{a}_p^\dag\tilde{b}_q^\dag.
 \end{equation}
The coefficient matrix $V$ and the normalization constant $N$ are
the unknowns to be evaluated. Using
(\ref{eq:DiracFieldRepresntation1}),(\ref{eq:DiracFieldRepresntation3}),
and (\ref{eq:ModeRelation13}), the coefficient matrix is given by
\begin{equation}\label{eq:Vpq}
V=V^{(0)}+V^{(1)}+O(h^2)=V^{(1)}+O(h^2),
\end{equation}with
\begin{equation}\label{eq:Vpq_Acalig_relation}
V^{(1)}_{pq}=\mathcal{A}_{pq}^{(1)*}G_q=-\mathcal{A}_{qp}^{(1)}G_p^*
\end{equation}

The relation between Fock vacua in regions I and III given by
(\ref{eq:vacuRegion13})
yields \cite{ARLEEPHDThesis,PhysRevD.85.025012}
\begin{equation}
\big|0\big>=\bigl(1-\frac{1}{2}\sum_{p\ge0,
q<0}|V_{pq}|^2\bigr)|\tilde{0}\big>
+\sum_{p,q}V_{pq}|\tilde{1}_p\big>^+|\tilde{1}_q\big>^-
-\frac{1}{2}\sum_{p,q}\sum_{i,j}V_{pq}V_{ij}\phi_{p,i}\phi_{q,j}
|\tilde{1}_p\big>^+|\tilde{1}_i\big>^+|\tilde{1}_q\big>^-|\tilde{1}_j\big>^-
\!\!+O(h^3).\label{eq:vacuRegion13FinalForm}
\end{equation}
where
{\small$|\tilde{1}_p\big>^+:=\tilde{a}_p^\dag|\tilde{0}\big>^+$} and
{\small$|\tilde{1}_q\big>^-:=\tilde{b}_q^\dag|\tilde{0}\big>^-$}
represent the single-particle Fock states for modes $p\ge0$ and
$q<0$, respectively. The sign $\pm$ in the superscript denotes the
sign of the charge. Further, the term
{\small$\phi_{pi}:=1-\delta_{p,i}$} is introduced to incorporate the
Pauli-exclusion principle for the single particle states with same
charge sign. Also the ordering of the single-particle kets
corresponds to the ordering of the fermionic creation operators
rather than the fermionic modes \cite{ARLEEPHDThesis}. Similarly, the charged single particle states in region III are
\cite{ARLEEPHDThesis,PhysRevD.85.025012}
\begin{widetext}
\begin{subequations}
\begin{flalign}
\nonumber k>0:\quad |1_k\big>^+=&-\sum_{p,q}V_{pq}\mathcal{A}_{qk}^*|\tilde{1}_p\big>^+
+\sum_{m\ge0}\mathcal{A}_{mk}^*\bigl\{\bigl(1-\frac{1}{2}\sum_{p,q}|V_{pq}|^2\bigr)
|\tilde{1}_m\big>^++
\sum_{p,q}V_{pq}\phi_{pm}|\tilde{1}_m\big>^+|\tilde{1}_p\big>^+|\tilde{1}_q\big>^-
&\\ &-\frac{1}{2}\sum_{p,q;i,j}V_{pq}V_{ij}\phi_{pi}\phi_{pm}\phi_{mi}\phi_{qj}|\tilde{1}_m\big>^+|\tilde{1}_p\big>^+
|\tilde{1}_i\big>^+|\tilde{1}_q\big>^-|\tilde{1}_j\big>^-
\bigr\}+O(h^3),& \\*
\nonumber k<0:\quad |1_k\big>^-=&\sum_{p,q}V_{pq}\mathcal{A}_{pk}|\tilde{1}_q\big>^-
+\sum_{m<0}\mathcal{A}_{mk}\bigl\{\bigl(1-\frac{1}{2}\sum_{p,q}|V_{pq}|^2\bigr)|\tilde{1}_m\big>^-+
\sum_{p,q}V_{pq}\phi_{qm}|\tilde{1}_p\big>^+|\tilde{1}_q\big>^-|\tilde{1}_m\big>^-&\\*
&-\frac{1}{2}\sum_{p,q;i,j}V_{pq}V_{ij}\phi_{pi}\phi_{qm}\phi_{qj}\phi_{mj}|\tilde{1}_p\big>^+|\tilde{1}_i\big>^+
|\tilde{1}_q\big>^-|\tilde{1}_j\big>^-|\tilde{1}_m\big>^-
\bigr\}+O(h^3),&
\end{flalign}\label{eq:OneParticleRegion13OpenForm}
\end{subequations}
\end{widetext}
where the one particle states $|1_k\big>^\pm$ in region I are
\begin{subequations}
\begin{eqnarray}
 k\ge0: \qquad |1_k\big>^+&=&a_k^\dag|0\big>,\\
 k<0: \qquad |1_k\big>^-&=&b_k^\dag|0\big>.
\end{eqnarray}\label{eq:OneParticleRegion13}
\end{subequations}
\section{%
Quantum Fisher Information for two-mode
states}\label{sec:QFI_FortwoMode_vac_1PrtclStates}
Here we investigate the behavior of the QFI for a complete trip of a two mode entangled states  from region I to region III in the perturbative regime
to the order $h^2$ and study how the uniform acceleration affects the QFI of the evolved fermionic modes confined to the cavities.
\indent Although a single cavity can reveal the effect of
non-inrtial (accelerated) motion in time, as described in
\cite{PhysRevD.85.061701} and \cite{PhysRevD.85.025012}, the
entangled mode in two cavities where one cavity remains inertial
throughout while the other may have segments of inertial and
non-inertial motion may show richer effect of non-inrtial
(accelerated) motion. And a localization of non-uniform
(accelerated) motion in time can then be achieved by considering the
initial and final segments of the accelerated cavity to be inertial
\cite{PhysRevD.85.025012}. We therefore employed and followed the
argument proposed by \cite{PhysRevD.85.061701} and
\cite{PhysRevD.85.025012} to consider the effects of non-uniform
motion on the precision of parameter estimation. In addition, we
have also considered a single cavity scenario for QFI analysis as an
extension.

 \subsection{%
 Evaluation of Perturbed Eigenvalues and Eigenvectors}\label{subsect:EigValuesAndVectors}
 We consider a bipartite two qubit pure state parameterized by parameter $\theta$
in region I. The state consists of two Dirac field modes where one of the modes is confined to Alice's cavity and the other is confined to Rob's cavity.
 The initial parameterized state is
\begin{equation}\label{eq:FermionicStateinitial_01}
\big|\psi^{\pm}_{init}\big>_\nu=\cos\theta \big|0\big>_A\big|0\big>_R\pm\sin\theta \big|1_m\big>_A^\mu\big|1_k\big>_R^\nu,
\end{equation}
where the subscripts A and R refer to the cavity of Alice and Bob,
respectively. In this setting, the initial state can be obtained by
a unitary operation which entangles the two confined modes of
cavities of Alice and Bob. However the rate of entanglement, which
is controlled by the parameter $\theta$, is unknown. In other words,
the parameter $\theta$ indicates the initial entanglement of the two
confined modes, and is not known in advance. However, the initially
chosen entangled state may provide some prior knowledge about the
range of the parameter but not the actual value. The superscripts
$\mu$ and $\nu$ indicate whether the mode has positive or negative
frequency, so that $\mu\left(\nu\right) = +$ for $m
\left(k\right)\ge 0$ and $\mu\left(\nu\right) = -$ for $m
\left(k\right)\le 0$. Following the procedure given in
\cite{ARLEEPHDThesis,PhysRevD.85.025012}, the initial state
(\ref{eq:FermionicStateinitial_01}) is represented by the
two-particle basis of the two mode Hilbert space with one excitation
for each of the modes $m$ and $k$ in Alice's  and Rob's cavities,
respectively. The corresponding density matrix in the region I is
written as
\begin{eqnarray}
\rho_\nu^{\pm}=\cos^2\theta\big|0\big>_A\big<0\big|\otimes\big|0\big>_R\big<0\big|
+\sin^2\theta\big|1_m\big>^{\mu\mu}_A\big<1_m\big|\otimes\big|1_k\big>^{\nu\nu}_R\big<1_k\big|+\big(\sin\theta\cos\theta\big|0\big>^\mu_A\big<1_m\big|\otimes\big|0\big>^{\nu}_R \big<1_k\big|+h.c\big).\label{eq:FermionicDensityMatrixInitial}
\end{eqnarray}
It should be noted that all the modes, except the reference mode in the Rob's cavity are related to the environment.
Therefore, a partial trace is taken over all of the Rob's cavity modes except the reference mode $k$.
By exploiting the unitarity of the perturbed Bogoliubov transformation (\ref{eq:TotalMatrixCompsition})  up to the second order perturbation
and using the inside out partial tracing approach \cite{PhysRevD.85.025012},
the reduced density matrix in the region
III is expressed as
\begin{widetext}
\begin{flalign}
\nonumber \textup{Tr}_{\neg k}\rho_\nu^{\pm}\equiv&\rho_{\nu,k}^\pm=\cos^2\theta\big|0\big>_A\big<0\big|\otimes\Big\{\left(1-f_k^{-\nu}h^2\right)\big|\tilde{0}\big>_{III}\big<\tilde{0}\big|
+f_k^{-\nu}h^2\big|\tilde{1}_k\big>_{III}\big<\tilde{1}_k\big|\Big\}
+\sin\theta\cos\theta\Big\{\pm\left(G_k+\mathcal{A}_{kk}^{\left(2\right)}h^2\right)&\\*
&\times\big|0\big>^\mu_A\big<1_m\big|\otimes\big|\tilde{0}\big>^{\nu}_{III} \big<\tilde{1}_k\big|\pm h.c\Big\}
+\sin^2\theta\big|1_m\big>^{\mu\mu}_A\big<1_m\big|\otimes\Big\{\left(1-f_k^{\nu}h^2\right)\big|\tilde{1}_k\big>^{\nu\nu}_{III}\big<\tilde{1}_k\big|+
f_k^{\nu}h^2\big|\tilde{0}\big>_{III}\big<\tilde{0}\big|\Big\}.&\label{eq:DensityMat_Finalized}
\end{flalign}
\end{widetext}

where $f_k^{\nu}$ and $f_k^{-\nu}$ are defined as
\begin{subequations}
\begin{eqnarray}
\nu>0: \qquad f_k^{\nu}&=&\sum_{p\ge0}|\mathcal{A}_{pk}^{(1)}|^2,\\*
\nu<0: \qquad f_k^{\nu}&=&\sum_{q<0}|\mathcal{A}_{qk}^{(1)}|^2.
\end{eqnarray}\label{eq:fk_PLUSMINUS}
\end{subequations}
The density matrix can be re-written as
\begin{equation}
\rho_{\nu,k}^\pm=\rho_{\nu,k}^{\pm\left(0\right)}+\rho_{\nu,k}^{\pm\left(2\right)}h^2,
\end{equation}
where $\rho_{\nu,k}^{\pm\left(0\right)}$ and
$\rho_{\nu,k}^{\pm\left(2\right)}$ denote the unperturbed and
perturbed matrix components, respectively. In order to evaluate the
QFI given by (\ref{eq:QauntumFIDesiredForm}) for the density matrix
in (\ref{eq:DensityMat_Finalized}), we compute the non-zero
eigenvalues and the corresponding eigenvectors.
The eigenvalues of the unperturbed part of the evolved density matrix $ \rho_{\nu,k}^{\pm(0)} $ are $\{p_i^{(0)}\} = \{ 1, 0, 0, 0\} $.  Note that the eigenvalues $1$ and $0$ denote the non-degenerate and degenerate case, respectively. We compute the second order corrections to the non-degenerate unperturbed eigenvalue $ p_1^{(0)} = 1 $ using standard perturbation procedure as prescribed in \cite{bransden2000quantum,reed1975fourier}. However, in case of the triply degenerate eigenvalue $ p_{2,3,4}^{(0)} = 0 $, the standard perturbation method is not valid and is needed to be replaced by the degenerate case. The second order corrections to the degenerate eigenvalue can be obtained by finding the eigenvalues of its degenerate subspace matrix $ M $ as described in \cite{bransden2000quantum}. Consequently, the eigenvalues of the perturbed density matrix are obtained as
\begin{eqnarray}
\textup{EigenVal}\left(\rho_{\nu,k}^{\pm}\right)=\left\{p_i\right\}=\Bigl\{1\!-\!\left(\cos^2\theta
f_k^{-\nu}+ \sin^2\theta f_k^{\nu} \right)h^2,\cos^2\theta f_k^{-\nu}h^2,\sin^2\theta f_k^{\nu}h^2,0\Bigr\}.\label{eq:perturbed_EigValues}
\end{eqnarray}
 Since the trace of the perturbed density matrix is $tr\left(\rho_{\nu,k}^{\pm}\right)=1$ and all the eigenvalues are non-negative,
 $\rho_{\nu,k}^{\pm}$ satisfies the density matrix representation. Again, using perturbation theory, the
normalized eigenvectors corresponding to the non-zero eigenvalues of
$\rho_{\nu,k}^\pm$ are
\begin{subequations}
\begin{flalign}
|\Phi_1\big>=&\frac{1}{\sqrt{N}}\Bigl\{G_k^{\left(\nu_*\right)}\left(\cos\theta-\alpha\sin\theta\right), 0, 0, \bigl(\sin\theta\!+\!\alpha\cos\theta\bigr)\Bigr\} ,&\label{eq:perturbed_EigV1}\\*
|\Phi_2\big>=&\bigl\{0, 1, 0, 0\bigr\},&\label{eq:perturbed_EigV2}\\*
|\Phi_3\big>=&\bigl\{0, 0, 1, 0\bigr\},&\label{eq:perturbed_EigV3}
\end{flalign}\label{eq:pert_EigVectors}
\end{subequations}
where $\alpha$ and the normalization constant $N$ are found to be
\begin{eqnarray}
\alpha&:=&\!\sin\theta \cos\theta\left(\!\frac{f_k^{-\nu}-f_k^{\nu}}{2}+i\textup{Im}\left(G_k\bar{\mathcal{A}}_{kk}^{\left(2\right)}\!\right)\!\right)h^2,
\label{eq:alpha_val}\\*
 N&=&1+|\alpha|^2.\label{eq:NormalConstIdentities}
\end{eqnarray}
 \subsection{%
 Quantum Fisher Information with respect to the parameter $\theta$}\label{subsec:QFI_Theta_Hparameters}
Next, we use (\ref{eq:QauntumFIDesiredForm}) to compute the QFI of
the Dirac field evolved from the region I to the region III. Prior
to the evolution, the QFI for the Dirac field of the initial
bipartite system (\ref{eq:FermionicDensityMatrixInitial}) with
respect to
parameter $\theta$ is $\mathcal{F}_\theta=4$. After the evolution of the system to the region III, we now compute quantum and classical contribution separately.  By using Eq.~(\ref{eq:QFIPureState}), the contribution to the QFI due to $|\Phi_1\big>$ is
\begin{eqnarray}
\nonumber\mathcal{F}_{\theta,1}&=&\big<\Phi_1'\big|\Phi'_1\big>-\left|\big<\Phi_1\big|\Phi'_1\big>\right|^2 \\*&=&1+\cos2\theta\left(f^{-\nu}_k-f^{\nu}_k\right)h^2\!+\!O(h^3),\label{eq:FQ1_Finallysimplified}
\end{eqnarray}
 where we have used (\ref{eq:alpha_val}) and (\ref{eq:NormalConstIdentities}) to get
\begin{equation}\label{eq:FQ1_3termsSimplified}
\frac{\textup{Re}\bigl(\alpha'\bigr)}{N}=\cos2\theta\frac{f^{-\nu}_k-f^{\nu}_k}{2}h^2+O(h^3).
\end{equation}
Here, we observe that the other two states namely $\big|\Phi_2\big>$ and $\big|\Phi_3\big>$ are independent of the
parameter $\theta$. Therefore, the individual contributions $\mathcal{F}_{\theta,2}$ and $\mathcal{F}_{\theta,3}$  disappear. Following the same argument, the mixed
terms $\big<\Phi_m\big|\Phi'_n\big>$ disappear as well $\forall
m\neq n$. Consequently, the net quantum contribution, $F_\theta$, to
the total QFI (\ref{eq:QauntumFIDesiredForm}) is computed  as
\begin{flalign}
\nonumber F_\theta=&4p_1\mathcal{F}_{\theta,1}&\\*
\nonumber=&4\left(1-\cos^2\theta f^{-\nu}_k h^2-\sin^2\theta f^{\nu}_k h^2+O(h^3)\right)\left(1+\cos2\theta\left(f^{-\nu}_k-f^{\nu}_k\right)h^2+O(h^3)\right)&
\\* =&4\left(1-\left(\sin^2\theta f^{-\nu}_k+\cos^2\theta f^{\nu}_k \right)h^2\right)+O(h^3).&\label{eq:FQNet}
\end{flalign}
Working perturbatively in the similar fashion and using the following relations
\begin{subequations}
\begin{eqnarray}
\frac{\left(p'_1\right)^2}{p_1}&=&O\left(h^3\right),\\*
\frac{\left(p'_2\right)^2}{p_2}&=&4\sin^2\theta f_k^{-\nu}h^2+O\left(h^3\right),\\*
\frac{\left(p'_3\right)^2}{p_3}&=&4\cos^2\theta\left(f_k^{\nu}\right)h^2+O\left(h^3\right),
\end{eqnarray}\label{eq:FClassicalIndividualPi_s}\
\end{subequations}
the classical contribution, $F_c$, to the QFI (\ref{eq:QauntumFIDesiredForm}) is given by
\begin{eqnarray}
\nonumber F_c&=&\sum_{i=1}^{3}\frac{\left(p'_i\right)^2}{p_i}\\*
&=&4\left(\sin^2\theta f_k^{-\nu}+\cos^2\theta f_k^{\nu}\right)h^2.\label{eq:FClassical}
\end{eqnarray}
Finally, using (\ref{eq:FQNet}) and (\ref{eq:FClassical}), the QFI
(\ref{eq:QauntumFIDesiredForm}) of the Dirac field modes confined to
Alice's and Rob's cavities becomes
\begin{equation}\label{eq:QauntumFIDesiredFormFinallySolved}
\mathcal{F}_\theta=F_\theta+F_c=4,
\end{equation}
where Rob's field mode has evolved as discussed before. It is worth
noticing that to the order $h^2$, the QFI of the bipartite cavities
system remains invariant even after the evolution of field mode in
Rob's cavity. In contrast to QFI, the entanglement of the maximally
entangled bipartite system $\bigl(\theta=\frac{\pi}{4}$ in
Eq.~$(\ref{eq:DensityMat_Finalized})\bigr)$ was found to have
periodic degradation under the same evolution process
\cite{PhysRevD.85.025012}. These results show that the entanglement
degradation of the maximally entangled state does not cause any loss
in the precision of the estimate of the parameter $\theta$ in the
perturbative regime to the order $h^2$.\par Furthermore, the
invariance of $\mathcal{F}_\theta$ through the inertial and
non-inertial segments of motion is highly nontrivial as for an
arbitrary single qubit state, Zhong et al. \cite{PhysRevA.87.022337}
have indicated that $\mathcal{F}_\theta$ remains unchanged only
through phase-damping channel. The QFI with respect to parameter
$\theta$ for scalar and Dirac field modes, which are not confined to
the cavities, under Unruh-Hawking effect \cite{Yao2014}, yields the
same result as computed
in our case.
\par It is also important to note that to compute the QFI, and in turn the Cramer-Rao bound, the process of inertial and non-inertial motion is repeated many times and there is a possibility that the perturbation in the motion may go beyond the control of the observer. In order to circumvent this situation, we can resort to the unitarity of the perturbed Bogoliubov transformation up to the second order perturbation. The unitarity of the perturbed Bogoliubov transformation ensures the positive definiteness of the evolved density matrix ($\rho\ge0$) and its normalization ($\textup{Tr}\rho=1$). Therefore, in each run,  we can consider the evolved state for the evaluation of the QFI if it satisfies the density matrix definition up to the second order perturbation and discard it otherwise.
\subsection{%
Quantum Fisher Information distribution over
subsystems}\label{subsec:QFIDISTRIBUTION} Lu et al.
\cite{PhysRevA.86.022342} provided an elegant hierarchical analysis
for the QFI of the composite system with respect to its constituent
subsystems for a variety of measurement settings. In a similar
fashion, we study the effect of accelerated motion of Rob's cavity
on the distribution of the QFI, $\mathcal{F}_\theta$, over constituent
subsystems. For the inertial fermionic cavity of Alice, it is
straightforward to show that $\mathcal{F}_\theta^A=4$.
\begin{figure}[h]
\centering{\includegraphics[height=2.0in,width=3.2in]{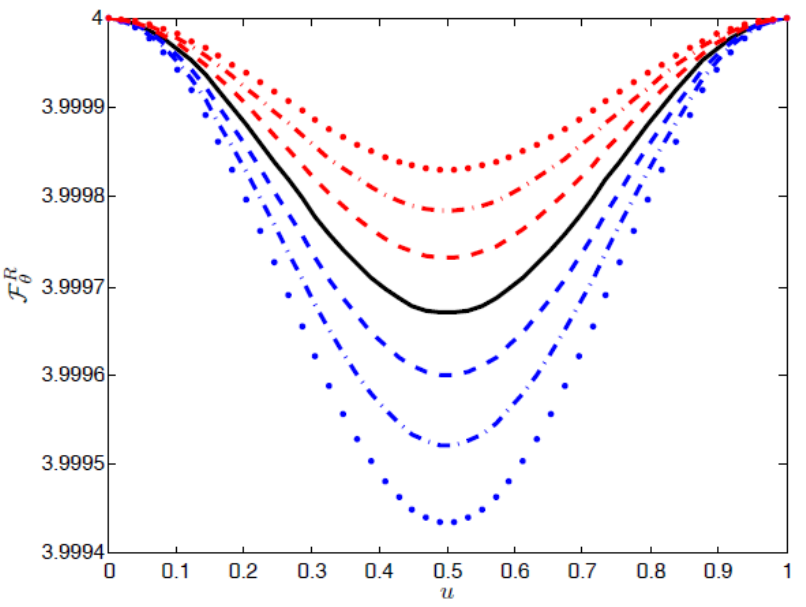}} \caption{The plot
shows the variation of QFI $\mathcal{F}_\theta^R; \theta=\pi/4$ over
Rob's cavity in region III as  a function of
$u:=\frac{1}{2}\eta_1/\ln(b/a)$ with $h=0.01$. The solid curve
(black) is for $s=0$ with $k=\pm1$. The dashed, dash-dotted and
dotted curves are, respectively, for $s=\frac{1}{4}$,
$\frac{1}{2}$,$\frac{3}{4}$, in k=1 (blue) above the solid curve and
k=-1 (red) below the solid curve.}\label{fig:fig1}
\end{figure}

Now let us consider the case for the fermionic mode in Rob's cavity in region III.
In this case, quantum part of the QFI, $F_\theta=0$.
Thus the only contribution to QFI, $\mathcal{F}_\theta^R$, is due to
its classical part, $F_c$.
Working pertubatively, as a result, we have
\begin{subequations}
\begin{flalign}
\nonumber\frac{\left(p_1'\right)^2}{p_1}=&\frac{\sin^22\theta\left(1-f_kh^2\right)^2}
{\cos^2\theta+\left(-\cos^2\theta f_k^{-\nu}+\sin^2\theta f_k^{\nu}\right)h^2}&\\*
=&4\sin^2\theta\!\left[\!1\!+\!\left(f_k^{-\nu}\!-\!f_k^{\nu}\tan^2\theta-2f_k\right)h^2\right],&\\*
\frac{\left(p_2'\right)^2}{p_2} =&4\cos^2\theta\!\left[1+\left(f_k^{\nu}\!-\! f_k^{-\nu}\cot^2\theta\!-\!2f_k\right)h^2\right],&
\end{flalign}
\end{subequations}
where $f_k^+$, $f_k^-$ given by (\ref{eq:fk_PLUSMINUS}) can be re-written as
\begin{subequations}
\begin{eqnarray}
f_k^{+}&=&\sum_{p\ge0}^{\infty}\left|E_1^{k-p}\right|^2|A_{pk}^{(1)}|^2,\\*
f_k^{-}&=&\sum_{q<0}^{\infty}\left|E_1^{k-q}\right|^2|A_{qk}^{(1)}|^2,\\*
f_k:&=&f_k^{+}+f_k^{-}=\sum_{p=-\infty}^{\infty}|\mathcal{A}_{pk}^{(1)}|^2,
\end{eqnarray}\label{eq:fks_All}
\end{subequations}
with
\begin{equation}
E1:=\exp\left(\frac{i\pi\eta_1}{\ln\left(b/a\right)}\right).
\end{equation}
The QFI $\mathcal{F}_\theta^R$ finally reads
\begin{eqnarray}\label{eq:QFI_ReducedRobState}
\displaystyle\mathcal{F}_\theta^R&=&4-4\frac{\left(\sin^2\theta f_k^{\nu}+\cos^2\theta f_k^{-\nu}\right)}{\sin^2\theta\cos^2\theta}h^2.
\\ \nonumber&&
\end{eqnarray}
Since by definition (\ref{eq:fks_All}), the relations $f_k^\nu$ and $f_k^{-\nu}$
are periodic and non-negative, the QFI over Rob's
subsystem is non-negative and is less than or equal to 4 with
$0<\theta< \pi /2 $ for the perturbative regime $|k|h\ll1$. Plots
for parameter $\theta=\pi/4$ with $k=\pm1$ are shown in
Fig.~\ref{fig:fig1}. Consequently,  the QFI of the bipartite
composite system, $\mathcal{F}_\theta$, is sub-additive.  Further, it can be noticed that the QFI shows periodic degradation over Rob's subsystem depending upon the duration of non-inertial motion of Rob' cavity.
However, following the procedure described in \cite{PhysRevD.85.025012}, this periodic degradation can be compensated by tuning the intervals of the inertial and non-inertial motion of the Rob's cavity.
\par As an extension to the QFI computation of the two mode state, we evaluate the QFI for a single fermionic mode restricted to one cavity and compare its variation with that of  Rob's cavity subsystem.
The same setting of inertial and non-inertial segments of motion is
considered for the single cavity with the exception that the
reference (inertial) cavity is not employed in this case. The
parameter $\theta$ is now encoded in the single fermionic cavity
mode and the initial state in region I is, therefore, written as
\begin{equation}\label{eq:SingleMode_Initial_Wavefcn}
\big|\psi_{init}\big>=\cos\theta\big|0\big>+\sin\theta\big|1_k\big>^\nu.
\end{equation}
The density matrix representation is given by
\begin{equation}\label{eq:SingleMode_Initial_DenMatrx}
\rho_{\nu}^S=\cos^2\theta\big|0\big>\big<0\big|
+\sin^2\theta\big|1_k\big>^{\nu\nu}\big<1\big|
+\sin\theta\cos\theta\left(\big|0\big>^\nu\big<1\big|+\big|1\big>^\nu\big<0\big|\right).
\end{equation}
Using the inside out partial tracing approach \cite{PhysRevD.85.025012} and by invoking the unitarity of the perturbed Bogoliubov transformation (\ref{eq:TotalMatrixCompsition}), the reduced density matrix in region III is expressed as
\begin{widetext}
\begin{eqnarray}
\nonumber\rho_{\nu,k}^S\equiv\textup{Tr}_{\neg{k}}\rho_\nu^S &=&
\left(\cos^2\theta(1-f_k^{-\nu}h^2)+\sin^2\theta f_k^{\nu}h^2\right)\big|\tilde{0}\big>\big<\tilde{0}\big|
+\left(\sin^2\theta(1-f_k^{\nu}h^2)+\cos^2\theta f_k^{-\nu}h^2\right)\big|\tilde{1}_k\big>^{\nu\nu}\big<\tilde{1}_k\big|\\
&&+\sin\theta\cos\theta\left(\left(G_k+\mathcal{A}_{kk}^{(2)}\right)
\big|\tilde{0}\big>^{\nu}\big<\tilde{1}_k\big|+h.c\right).\label{eq:Evolved_DenMat_SingleCavity}
\end{eqnarray}
\end{widetext}
Since both the eigenvalues $\{p_i\}=\{1,0\}$ of the unperturbed part are non-degenerate, therefore the second order correction to these eigenvalues can be obtained using standard perturbation theory as described in \cite{bransden2000quantum}. The non-zero eigenvalues of the evolved state are given by
\begin{equation}
\{p_i\}=\left\{1-\left(\cos^4\theta f_k^{-\nu}+\sin^4\theta f_k^{\nu}\right)h^2,
\left(\cos^4\theta f_k^{-\nu}+\sin^4\theta f_k^{\nu}\right)h^2\right\}.
\end{equation}
The corresponding normalized eigenvectors are
\begin{subequations}
\begin{eqnarray}
\big|\Phi_1\big>&=&\frac{1}{\sqrt{N}}
\left\{
\begin{array}{cc}
  G_k\left(\cos\theta-\alpha\sin\theta\right) & \left(\sin\theta+\alpha\cos\theta\right)
\end{array}
\right\},\\
\big|\Phi_2\big>&=&\frac{1}{\sqrt{N}}
\left\{
\begin{array}{cc}
 -G_k\left(\sin\theta+\bar{\alpha}\cos\theta\right) & \left(\cos\theta-\bar{\alpha}\sin\theta\right)
\end{array}
\right\},
\end{eqnarray}
\end{subequations}
where $N=1+|\alpha|^2$ is as usual the normalization constant and $\alpha$ is defined as
\begin{equation}
\alpha=\sin\theta\cos\theta
\biggl[\frac{1}{2}\cos2\theta f_k
+\left(f_k^{-\nu}-f_k^{\nu}\right)
+i\textup{Im}\left(G_k\bar{\mathcal{A}}_{kk}^{(2)}\right)
\biggr]h^2.
\end{equation}
The classical contribution to the QFI is given by
\begin{eqnarray}
\nonumber F_c&=&\sum_i \frac{(p_i')^2}{p_i}\\
&=&16\cos^2\theta\sin^2\theta
\frac{\left(\cos^2\theta f_k^{-\nu}-\sin^2\theta f_k^{\nu}\right)^2}
{\left(\cos^4\theta f_k^{-\nu}+\sin^4\theta f_k^{\nu}\right)}h^2+O(h^4).
\label{eq:FClassical_SingleCavityMode}
\end{eqnarray}
The second term involved in the QFI, namely the quantum contribution due to the individual pure states, is computed as follows
\begin{eqnarray}
\nonumber F_{\theta;i}&=&4\sum_i
p_i\left(\big<\Phi'_i|\Phi'_i\big>-\big|\big<\Phi_i|\Phi'_i\big>\big|^2\right)\\
&=&4+4\left\{
\cos4\theta f_k
+2\cos2\theta\left(f_k^{-\nu}-f_k^{\nu}\right)
\right\}h^2.\label{eq:FQuantum_pure_SingleCavityMode}
\end{eqnarray}
 Further, the quantum contribution due to the mixture of pure states can be expressed as
\begin{eqnarray}
\nonumber F_{\theta;ij}&=&8\sum_{i\neq j}
\frac{p_ip_j}{p_i+p_j}\big|\big<\Phi_i|\Phi'_j\big>\big|^2\\
&=&16\left(\cos^4\theta f_k^{-\nu}+\sin^4\theta f_k^{\nu}\right)h^2.
\label{eq:FQuantum_mixed_SingleCavityMode}
\end{eqnarray}
Finally, using the relations (\ref{eq:FClassical_SingleCavityMode})-(\ref{eq:FQuantum_mixed_SingleCavityMode}), the QFI for the single fermionic mode confined to the single cavity can be expressed as
\begin{equation}\label{eq:QFI_SingleCavityMode}
\mathcal{F}_\theta^S=F_c+F_{\theta;i}-F_{\theta;ij}.
\end{equation}
\begin{figure}[h]
\centering{\includegraphics[height=2.0in,width=3.2in]{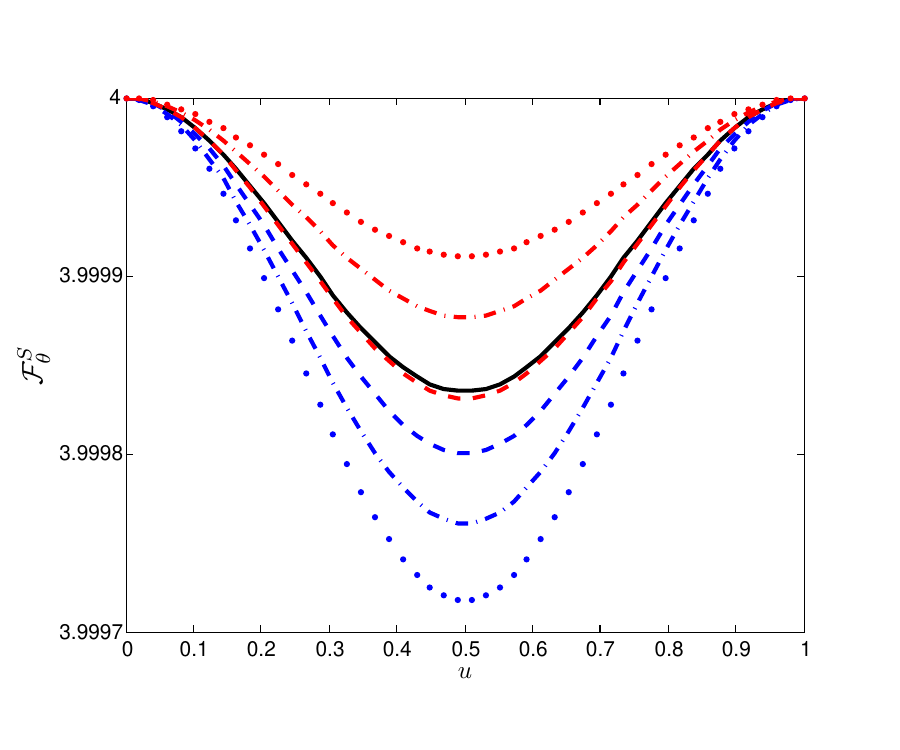}}
\caption{The plots
show the variation of QFI $\mathcal{F}_\theta^S$; $\theta=\pi/4$, of a single fermionic mode within one cavity in region III as a function of $u:=\frac{1}{2}\eta_1/\ln(b/a)$ with $h=0.01$. The solid curve
(black) is for $s=0$ with $k=\pm1$. The dashed, dash-dotted and
dotted curves are, respectively, for $s=\frac{1}{4}$,
$\frac{1}{2}$,$\frac{3}{4}$, in k=1 (blue) and
k=-1 (red).}\label{fig:fig2}
\end{figure}
As shown in Fig.~\ref{fig:fig2}, the QFI behavior in the case of single Dirac field mode confined to the single cavity is similar to that of the Rob's cavity subsystem (Fig.~\ref{fig:fig1}).
However, it is different from the QFI variation (\ref{eq:QauntumFIDesiredFormFinallySolved}) of the bipartite or composite system which remains invariant
and shows no degradation up to the second order corrections.
\section{%
Quantum Fisher Information for the Werner
state}\label{sec:QFI_WernerState} We have studied the behavior of
Fisher information for the entangled confined modes of cavities of
Alice and Bob, where the amount of entanglement is controlled by the
parameter $\theta$. We have found the invariance of Fisher
information regardless of the parameter $\theta$ up to the order
$h^2$. Therefore it is natural to ask whether the degrade of
entanglement does not affect on the Fisher information. Here, we
study the dynamics of QFI for the Werner state, which is a depolaring state of the $\theta$ parameterized two-qubit pure entangled state. For this purpose, we consider the initial two qubit
Werner state in region I given by
\begin{eqnarray}
\rho_\nu=r\left(\cos\theta\big|0\big>\big|0\big>+\sin\theta\big|1_m\big>^{\mu}\big|1_k\big>^{\nu}\right)
\left(\cos\theta\big<0\big|\big<0\big|+\sin\theta^{\mu}\big<1_m\big|^{\nu}\big<\big|1_k\big|\right)+\frac{1-r}{4}\mathbb{I},
\end{eqnarray}\label{eq:WernerDensityMatrixInitial}
where the parameter $r$ indicates the mixedness of the pure entangled
two qubit state and the maximally mixed bipartite state. Considering
the perturbed evolution of the state in Rob's cavity from region I
to region III in the similar fashion as discussed earlier,  we
transform the density matrix part confined to Rob's cavity in terms
of Rob's Region (III) basis to the order $h^2$ with the help of
(\ref{eq:vacuRegion13FinalForm}) and
(\ref{eq:OneParticleRegion13OpenForm}). Afterwards, by exploiting the unitarity of the Bogoliubov transformation (\ref{eq:TotalMatrixCompsition}) and applying the partial trace as described in \cite{PhysRevD.85.025012}, the reduced density matrix in the region III is expressed as
\begin{widetext}
\begin{eqnarray}
\nonumber
\rho_{_W;\nu,k,}&=&r\Bigl[\cos^2\theta\big|0\big>_A\big<0\big|\otimes\Big\{\left(1-f_k^{-\nu}h^2\right)\big|\tilde{0}\big>_{III}\big<\tilde{0}\big|
+f_k^{-\nu}h^2\big|\tilde{1}_k\big>_{III}\big<\tilde{1}_k\big|\Big\}
+\sin\theta\cos\theta\Big\{\left(G_k+\mathcal{A}_{kk}^{\left(2\right)}h^2\right)\\*
\nonumber&&\times\big|0\big>^\mu_A\big<1_m\big|\otimes\big|\tilde{0}\big>^{\nu}_{III}
\big<\tilde{1}_k\big|\pm h.c\Big\}
+\sin^2\theta\big|1_m\big>^{\mu\mu}_A\big<1_m\big|\otimes\Big\{\left(1-f_k^{\nu}h^2\right)\big|\tilde{1}_k\big>^{\nu\nu}_{III}\big<\tilde{1}_k\big|+
f_k^{\nu}h^2\big|\tilde{0}\big>_{III}\big<\tilde{0}\big|\Big\}\Bigr]\\*
&&+r^c\Bigl[\left(\big|0\big>_A\big<0\big|+\big|1_m\big>^{\mu\mu}_A\big<1_m\big|\right)
\otimes\Big\{\left(1+g_k^\nu
h^2\right)\big|\tilde{0}\big>_{III}\big<\tilde{0}\big|
+\left(1+g_k^{-\nu }
h^2\right)\big|\tilde{1}_k\big>_{III}\big<\tilde{1}_k\big|\Big\}
\Bigr],\label{eq:WernerDensityMat_Finalized}
\end{eqnarray}
\end{widetext}
where $r^c$ and $g_k^{\pm\nu}$ are defined as
\begin{subequations}
\begin{eqnarray}
r^c&=&\frac{1-r}{4},\\* g_k^{\pm\nu}&=&f^{\pm\nu}_k-f^{\mp\nu}_k.
\end{eqnarray}\label{eq:Fc_gkAbbrev}
\end{subequations}
Thus the perturbed density matrix can be expressed in the compact form as
\begin{equation}\label{eq:WernerDensityMat_PerturbedForm}
\rho_{_W;\nu,k}=\rho_{_W;\nu,k}^{(0)}+\rho_{_W;\nu,k}^{(2)}h^2.
\end{equation}
The unperturbed part of the density matrix, $\rho_{_W;\nu,k}^{(0)}$,  admits the same set of the eigenvectors as obtained for the two mode pure state in Sec.~\ref{sec:QFI_FortwoMode_vac_1PrtclStates} corresponding to the respective unperturbed eigenvalues
\begin{equation}\label{eq:unpert_EigValuesWerner}
\textup{EigenVal}\left(\rho_{_W;\nu,k}^{\left(0\right)}\right)=\{r+r^c,r^c,r^c,r^c\}.
\end{equation}
Two explicit cases arise for $r=0$ and $r=1$. For $r=0$, the
unperturbed density matrix represents the maximally mixed state,
with standard basis and degenerate eigenvalue of
$p_{1,2,3,4}^{(0)}=1$. In this case, the QFI with respect to the
parameter $\theta$ yields a trivial result, $\mathcal{F}_\theta=0$.
For $r=1$, the situation is exactly the same as described in the
Sec. \ref{sec:QFI_FortwoMode_vac_1PrtclStates}. Therefore we
restrict the value of $r$, in the open interval $0<r<1$. \par It can be noted that, the eigenvalues $r+r^c$ and $r^c$  of the unperturbed part of the density matrix denote the non-degenerate and triply degenerate cases, respectively.  Therefore, following the perturbative procedure prescribed in \cite{reed1975fourier,bransden2000quantum} and used in the previous section for the non-degenerate and degenerate cases, the eigenvalues of the reduced density matrix $\rho_{_W;\nu,k}$ in the region III are
\begin{flalign}
\nonumber
\nonumber\{p_i\}=\Bigl\{&
r+r^c-r\left(\cos^2\theta f_k^{-\nu}+\sin^2\theta
f_k^{\nu}\right)h^2+r^c\left(\cos^2\theta
g_k^\nu+\sin^2\theta g_k^{-\nu}\right)h^2,
r^c+r^c\left(\sin^2\theta g_k^\nu+\cos^2\theta
g_k^{-\nu}\right)h^2,&\\* &r^c+r^cg_k^{-\nu}h^2+r
f_k^{-\nu}\cos^2\theta h^2, r^c+r^cg_k^{\nu}h^2+r
f_k^{\nu}\sin^2\theta h^2\Bigr\}.&\label{eq:pert_EigValuesWerner}
\end{flalign}
From (\ref{eq:pert_EigValuesWerner}), it can be seen that the
perturbed density matrix satisfies the density matrix conditions,
$p_i\ge0; \forall i$ and $\sum_i p_i=1$. Furthermore, the classical
contribution, to the QFI, $F_c$,  vanishes for $0<r<1$. This is due to the fact that for each eigenvalue, $p_i$, the expression $(\partial_\theta
p_i)^2/p_i$ has the leading term of order $h^4$. Hence,
$F_c=0+O(h^4)$. However, it is important to note that for the special
case of $r=1$, the classical contribution $F_c$ is non-vanishing and
is given by (\ref{eq:FClassical}). Next, the corresponding perturbed
eigenvectors are
 \begin{figure}[b]
 \centering \includegraphics[height=2.0in,width=3.0in]{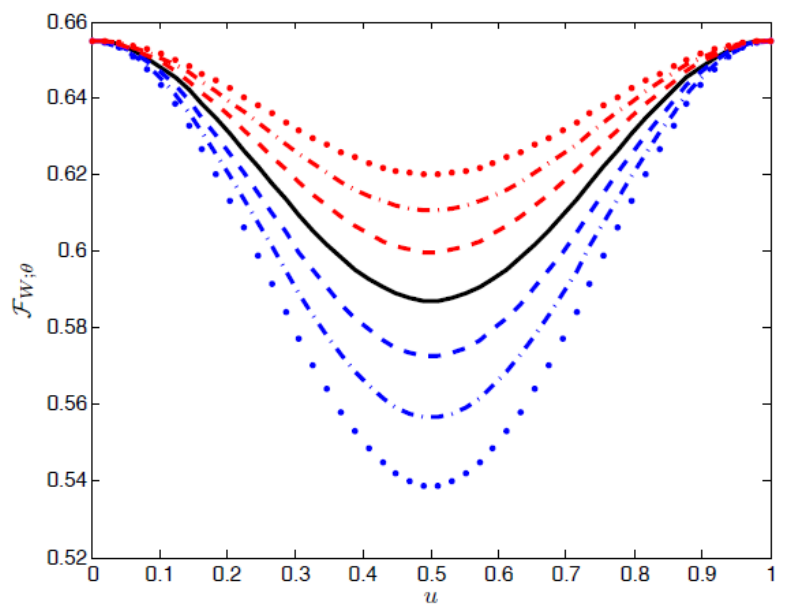}\\  \caption{The plot shows the variation of QFI
  $\mathcal{F}_{_W;\theta}$
at $\theta=\pi/4$ for  bipartite Werner  state in region III as  a
function of $u:=\frac{1}{2}\eta_1/\ln(b/a)$ with fixed value:
$h=0.01$, $r=1/3$. The solid curve (black) is for $s=0$ with
$k=\pm1$. The dashed, dash-dotted and dotted curves are,
respectively, for $s=\frac{1}{4}$, $\frac{1}{2}$,$\frac{3}{4}$, in
k=1 (blue) above the solid curve and k=-1 (red) below the solid
curve.}\label{fig:fig3}
\end{figure}

\begin{subequations}
\begin{flalign}
|\Phi_1\big>=&\frac{1}{\sqrt{N}}\left\{
                \begin{array}{cccc}
                  G_k\left(\cos\theta-\frac{\beta}{r}\sin\theta\right) & 0 & 0 & \sin\theta\!+\!\frac{\beta}{r}\cos\theta\\
                \end{array}
              \right\},
 &\label{eq:perturbed_EigV1Werner}\\*
 |\Phi_2\big>=&\frac{1}{\sqrt{N}}\!\! \left\{
                \begin{array}{cccc}
                \!\! -G_k\left(\sin\theta+\frac{\bar{\beta}}{r}\cos\theta\right) & 0 & 0 & \cos\theta\!-\!\frac{\bar{\beta}}{r}\sin\theta \\
                \end{array}
              \!\! \right\},
 &\label{eq:perturbed_EigV2Werner}\\*
 |\Phi_3\big>=&\left\{
                \begin{array}{cccc}
                 0 & 1 & 0 & 0 \\
                \end{array}
              \right\},
 &\label{eq:perturbed_EigV3Werner}\\*
  |\Phi_4\big>=&\left\{
                \begin{array}{cccc}
                 0 & 0 & 1 & 0 \\
                \end{array}
              \right\},&\label{eq:perturbed_EigV4Werner}
\end{flalign}\label{eq:pert_EigVectorsWerner}
\end{subequations}
where $N=1+|\beta|^2/r^2$ is the normalization constant and
$\beta$ is defined as
 \begin{equation}\label{eq:beta_val}
 \beta=\sin\theta\cos\theta\Bigl[\frac{f_k^{-\nu}-f_k^{\nu}}{2}+ir\textup{Im}(G_k\bar{\mathcal{A}}_{kk}^{(2)})\Bigr]h^2.
 \end{equation}
 The quantum contribution due to the individual quantum states is
\begin{equation}
F_{\theta;i}= 4\sum_ip_i\mathcal{F}_{\theta,i}=4\biggl[\frac{1+r}{2}+\Bigl\{\frac{1+r}{2r}\cos2\theta\bigl(f_k^{-\nu}-f_k^{\nu}\bigr)
-r\bigl(\cos^2\theta f_k^{-\nu}+\sin^2\theta f_k^{\nu}\bigr)
\Bigr\}h^2\biggr].\label{eq:QFIPureStateWerner}
\end{equation}
Next, we have the contribution to the QFI due to the mixture, which
is given by
\begin{flalign}
\nonumber F_{\theta;ij}=&8\sum_{i\neq j}\mathcal{F}_{\theta;ij}&\\*
\nonumber=&16\biggl[\frac{1+2r-3r^2}{8\left(1+r\right)}\Bigl\{1+\frac{\cos2\theta}{r}\left(f_k^{-\nu}-f_k^{\nu}\right)h^2\Bigr\}+\left(1-r\right)r\Bigl\{
\frac{1+3r}{4\left(1+r\right)^2}\left(\sin^2\theta
f_k^{\nu}+\cos^2\theta f_k^{-\nu}\right)\\*
&-\frac{1}{2\left(1+r\right)}\left(\sin^2\theta
f_k^{-\nu}+\cos^2\theta f_k^{\nu}\right)\Bigr\}h^2
\biggr].&\label{eq:QFIMixedStateWerner}
\end{flalign}
The QFI $\mathcal{F}_\theta$ of the Werner state,
(\ref{eq:WernerDensityMat_Finalized}), with $0<r<1$ can therefore be
expressed using (\ref{eq:QFIPureStateWerner}) and
(\ref{eq:QFIMixedStateWerner}) as
\begin{equation}\label{eq:QFIWerner_Btw01}
\mathcal{F}_\theta=F_\theta=F_{\theta;i}-F_{\theta;ij}.
\end{equation}
It can be noticed that for $r<1$, the classical contribution, $F_c$,
is zero while the quantum mixture, $F_{\theta;ij}$, is non-zero.
However, for $r=1$, the classical contribution is non-vanishing
while the mixture term disappears and we obtain the same result (\ref{eq:QauntumFIDesiredFormFinallySolved}) as a special case.
Unlike the pure two qubit state case earlier discussed in Sec.
\ref{sec:QFI_FortwoMode_vac_1PrtclStates}, it can be seen from
(\ref{eq:QFIWerner_Btw01}) that the QFI of the Werner state is affected
due to  the inertial and non-inertial segments of Rob's cavity motion
for mixing parameter $r\in\left(0,1\right)$. Furthermore, the QFI of the Werner state
exhibits periodic degradation for $r\in\left(0,1\right)$
as shown in Fig. \ref{fig:fig3} for the case of $\theta=\pi/4$ and $r=1/3$. This
periodic degradation, however, can be avoided by
fine-tuning the duration of inertial and non-inertial trajectories for Rob's cavity.\\
The effect of mixing parameter $r$ is also investigated and is shown
in Fig. \ref{fig:fig4}. 
\begin{figure}[h]
  \centering\includegraphics[height=2.0in,width=3.2in]{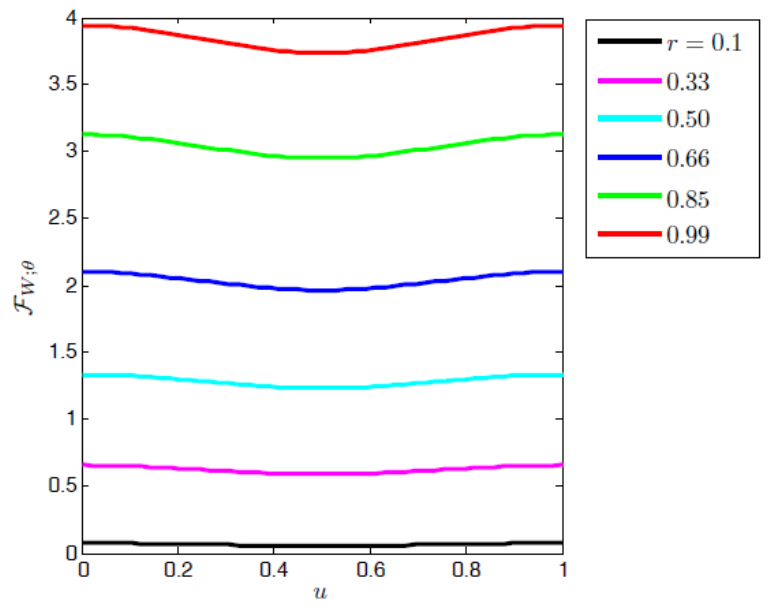}\\  \caption{The plot shows the variation of QFI
  $\mathcal{F}_{_W;\theta}$
at $\theta=\pi/4$ for bipartite Werner state in region III as  a
function of $u:=\frac{1}{2}\eta_1/\ln(b/a)$ with $h=0.01$. The
colored curves from bottom to top correspond to the values of
parameter $r=0.1, 0.33, 0.5, 0.66, 0.85, 0.99$, respectively. The
mode in Rob's cavity is kept fixed at $k=1$ with $s=0$ for all these
curves. }\label{fig:fig4}
\end{figure}
It is shown that as the parameter $r$ varies(in the region of $0<r<1$), the QFI of Werner state is affected by the inertial and non-inertial segments of motion of Rob's cavity.
\par The QFI distribution over subsystems of Alice's and Rob's cavity modes can be evaluated
using a procedure similar to that of Sec.
\ref{subsec:QFIDISTRIBUTION}. The QFI $\mathcal{F}_\theta^A$ over
Alice's cavity in region III is given by
\begin{equation}\label{QFI_ReducedAliceWernerState}
\mathcal{F}_\theta^A=\frac{4r^2\sin^22\theta}{1-r^2\cos^22\theta}.
\end{equation}
The QFI $\mathcal{F}_\theta^R$ over Rob's cavity in region III is given by
\begin{flalign}
\mathcal{F}_\theta^R=&\frac{4r^2\sin^22\theta}{1-r^2\cos^22\theta}\biggl[1-2f_kh^2
-\frac{r\cos2\theta}{1-r\cos^22\theta} \Bigl\{4r\bigl(\cos^2\theta f_k^{-\nu}\! \!\!-\!\sin^2\theta
f_k^{\nu}\bigr)\!+\!2(1\!-\!r)\bigl(f_k^{-\nu} \!\!-\!\!
f_k^{\nu}\bigr)\Bigr\}h^2
\!\biggr].&\label{QFI_ReducedRobsWernerState}
\end{flalign}
Since the QFI for each cavity mode is non-negative and has a maximum
value of 4, therefore the QFI distribution over subsystems is sub-additive.\\ \\
\section{%
Conclusion and Discussion}\label{sec:QFI_conclusion and discussion}

We investigated the effect of relativistic motion on the quantum
Fisher information of the $\left(1+1\right)$  Dirac field modes
confined to cavities. For this purpose, we considered a realistic
scheme for two cavities. In our scenario, the modes of a massless
Dirac field were confined to the two cavities with Dirichlet��s
boundary conditions, where one of the cavities remained inertial,
while the other underwent the segments of inertial and non-inertial
motion with uniform acceleration. The acceleration was assumed to be
very small and its effects were analyzed in a perturbative regime.
We considered a $\theta$ parameterized two-qubit pure entangled state
and a Werner state. In contrast to the degradation of entanglement
due to the relativistic motion between the cavities, the quantum
Fisher information of the pure composite system $\mathcal{F}_\theta$
with respect to parameter $\theta$ was found to be invariant under
the same conditions. However, in the case of the Werner state, 
which is a depolaring state of the $\theta$ parameterized two-qubit pure entangled state,
the quantum Fisher information displayed periodic degradation, due to
the inertial and non-inertial segments of motion. Furthermore, the quantum Fisher information over Rob's cavity showed periodic degradation behavior depending upon the parameter $\theta$
as well as the uniform acceleration for both the two qubit pure
state and Werner state.  The quantum Fisher information over Alice's
cavity remained invariant throughout the motion of Rob's cavity for
the two qubit pure state, whereas for the Werner state it was
affected by the mixing parameter of the Werner state. The subadditivity of the quantum Fisher information was held regardless of the pure composite system or the Werner state.
\section*{Acknowledgment}
Y.Kwon is supported by the Basic Science Research Program through
the National Research Foundation of Korea funded by the Ministry of
Education, Science and Technology (NRF2010-0025620 and NRF2015R1D1A1A01060795).



\begin{thebibliography}{99} 

\bibitem{Braunstein1996135} S. L. Braunstein, C. M. Caves, and G. J. Milburn, Annals of Physics \textbf{247}, 135 (1996).

\bibitem{PhysRevLett.72.3439} S. L. Braunstein,  and C. M. Caves,  Phys. Rev. Lett. \textbf{72}, 3439 (1994). 

\bibitem{1976quantum} C. W. Helstorm, \emph{Quantum detection and estimation theory}, Mathematics in Science and Engineering (Elsevier Science, 1976), ISBN 9780080956329. 

\bibitem{PhysRevLett.91.180403} S. Luo, Phys. Rev. Lett. \textbf{91}, 180403 (2003). 

\bibitem{PhysRevA.84.042121} Y. Watanabe, T. Sagawa, and M. Ueda, Phys. Rev. A \textbf{84}, 042121 (2011). 

\bibitem{Fisher1925} R. A. Fisher, in \emph{Mathematical Proceedings of the Cambridge Philosophical Society} (Cambridge Univ. Press, 1925), vol. 22, pp. 700-725. 

\bibitem{holevo2011probabilistic} A. S. Holevo, \emph{Probabilistic and statistical aspects of quantum theory}, Mathematics in Science and Engineering (Springer, 2011), ISBN 9788876423758. 

\bibitem{PhysRevD.14.2460} S. W. Hawking, Phys. Rev. D \textbf{14}, 2460 (1976). 

\bibitem{PhysRevD.14.870} W. G. Unruh, Phys. Rev. D \textbf{14}, 870 (1976). 

\bibitem{Wald94quantumfield} R. M. Wald, \emph{Quantum Field Theory in Curved Spacetime and Black Hole Thermodynamics}, (University of Chicago Press, Chicago, 1994). 


\bibitem{fabbri2005modeling} A. Fabbri, and J.Navarro-Salas, \emph{Modeling black hole evaporation}, (World Scientific, 2005). 


\bibitem{RevModPhys.76.93} A. Peres, and D. R. Terno, Rev. Mod. Phys. \textbf{76}, 93 (2004).

\bibitem{AlcingObDEpEnt-0264-9381-29-22-224001} P. M. Alsing, and I. Fuentes, Classical and Quantum Gravity \textbf{84}, 224001 (2012). 

\bibitem{PhysRevA.82.042332} D. E. Bruschi, J. Louko, E. Mart{\'i}n-Mart{\'i}nez, A. Dragan, and I. Fuentes, Phys. Rev. A \textbf{82}, 042332 (2010);
J. Chang and Y.Kwon,Phys. Rev. A \textbf{85},  032302  (2012); J. Chang and Y.Kwon, Int.J.Theo.Phys. \textbf{54}, 996 (2015) 

\bibitem{RevModPhys.80.787} L. C. B. Crispino, A. Higuchi, and G. E. A. Matsas, Rev. Mod. Phys. \textbf{80}, 787 (2008). 

\bibitem{PhysRevLett.95.120404} I. Fuentes-Schuller, and R. B. Mann, Phys. Rev. Lett. \textbf{95}, 120404 (2005);
 M. Montero and E. Martin-Martinez, Phys. Rev. A 83, 062323 (2011); M. Montero and E.
 Martin-Martinez, JHEP 07, 006 (2011); J.Chang and Y.Kwon, Phys. Rev. A \textbf{86}, 014302  (2012)

\bibitem{PhysRevD.85.061701} D. E. Bruschi, I. Fuentes, and J. Louko, Phys. Rev. D \textbf{85}, 061701 (2012).

\bibitem{PhysRevA.87.022338} N. Friis, A. R. Lee, and D. E. Bruschi, Phys. Rev. A \textbf{87}, 022338 (2013).

\bibitem{PhysRevD.85.025012} N. Friis, A. R. Lee, D. E. Bruschi, and J. Louko, Phys. Rev. D \textbf{85}, 025012 (2012).

\bibitem{PhysRevA.74.032326} P. M. Alsing, I. Fuentes-Schuller, R. B. Mann, and  T. E. Tessier, Phys. Rev. A \textbf{74}, 032326 (2006).

\bibitem{PhysRevA.80.052304} A. Datta, Phys. Rev. A \textbf{80}, 052304 (2009). 


\bibitem{PhysRevA.81.052120} J. Wang, J. Deng, and J. Jing, Phys. Rev. A \textbf{81}, 052120 (2010). 


\bibitem{PhysRevA.86.032108} E. G. Brown, K. Cormier, E. Mart{\'i}n-Mart{\'i}nez, and R. B. Mann, Phys. Rev. A \textbf{86}, 032108 (2012);
J. Wang, J. Jing, and H. Fan, Phys. Rev. D \textbf{90} 025032 (2014)

\bibitem{PhysRevLett.91.180404} P. M. Alsing, and G. J. Milburn, Phys. Rev. Lett. \textbf{91}, 180404 (2003).


\bibitem{PhysRevLett.105.151301} M. Aspachs, G. Adesso, and I. Fuentes, Phys. Rev. Lett. \textbf{105}, 151301 (2010). 

\bibitem{Yao2014} Y. Yao, X. Xiao, L. Ge, X.-g. Wang, and C.-p. Sun,  Phys. Rev. A \textbf{89}, 042336 (2014); N. Metwally, arXiv:1609.02092 

\bibitem{MehdiAhmadi2013} M. Ahmadi, D. E. Bruschi, C. Sab{\'i}n, G. Adesso, and I. Fuentes, Sci. Rep. \textbf{4}, 4996 (2014). 


\bibitem{BuresMR0236719} D. Bures, Trans. Amer. Math. Soc. \textbf{135}, 199 (1969). 

\bibitem{PARIS2009} M. G. A. Paris, Int. J. of Quantum Inform. \textbf{07},125 (2009). 

\bibitem{PhysRevA.88.043832} Y. M. Zhang, X. W. Li, W. Yang, and G. R. Jin,   Phys. Rev. A \textbf{88}, 043832 (2013). 



\bibitem{ARLEEPHDThesis} A. R. Lee, Ph.D. thesis,
 University of Nottingham, 2013 (arXiv:1309.4419). 


\bibitem{reed1975fourier} M. Reed, and B. Simon, \emph{Methods of Modern Mathematical Physics}, (Academic Press, New York, 1975). 




\bibitem{bransden2000quantum} B. H. Bransden, and C. J. Joachain,  \emph{Quantum mechanics}, (Prentice hall Harlow, 2000). 


\bibitem{PhysRevA.87.022337} W. Zhong, Z. Sun, J. Ma, X. Wang, and F. Nori, Phys. Rev. A \textbf{87}, 022337 (2013). 


\bibitem{PhysRevA.86.022342} X.- M. Lu, S. Luo, and C. H. Oh, Phys. Rev. A \textbf{86}, 022342 (2012). 

\bibitem{PhysRevLett.110.113602} N. Friis, A. R. Lee, K. Truong, C. Sab{\'i}n,  E. Solano, G. Johansson, and I. Fuentes, Phys. Rev. Lett. \textbf{110}, 113602 (2013).

\bibitem{PhysRevD.89.065028} M. Ahmadi, D. E. Bruschi, and I. Fuentes, Phys. Rev. D \textbf{89}, 065028 (2014).

\bibitem{PhysRevA.40.4277} Werner, R. F., Phys. Rev. A \textbf{40}, 4277 (1989).




\end{thebibliography}

\end{document}